# DNA-based Artificial Nanostructures:

# Fabrication, Properties, and Applications

Young Sun and Ching-Hwa Kiang[*]

Department of Physics & Astronomy, Rice University

6100 Main Street - MS61, Houston, TX 77005, USA

Phone: (713) 348-4130, Fax: (713) 348-4150, E-mail: chkiang@rice.edu

[*]To whom correspondence should be addressed: chkiang@rice.edu.

**Table of Content**

# 1. Introduction

The integration of nanotechnology with biology and bioengineering is producing many advances. The essence of nanotechnology is to produce and manipulate well-defined structures on the nanometer scale with high accuracy. Conventional technologies based on the "top-down" approaches, such as the photolithographyic method, are difficult to continue to scale down due to real physical limitations including size of atoms, wavelengths of radiation used for lithography, and interconnect schemes. While engineers and scientists have long aspired to controllably manipulate structures at the micrometer and nanometer scale, nature elegantly performs these tasks and assembles structures with great accuracy and high efficiency using specific biological molecules.

Biological molecules, such as DNA, have shown great potential in fabrication and construction of nanostructures and devices. DNA molecules can be used for the assembly of devices and computational elements, for the assembly of interconnects, or as the device element itself. There are several advantages to use DNA for these constructions. First, DNA is the molecule whose intermolecular interactions are the most readily programmed and reliably predicted: Docking experiments reduce to the simple rules that A pairs with T, and G pairs with C. Thus, the very properties that make DNA so effective as genetic material also make it an excellent molecule for programmed self-assembly. Second, DNA of arbitrary sequences is available by convenient solid support synthesis. The needs of the biotechnology industry have also led to reliable chemistries to produce modifications, such as biotin groups, fluorescent labels, and linking functions. Third, DNA can be manipulated and modified by a large battery of enzymes that include DNA ligases, restriction endonucleases, kinases, and exonucleases.

On the other hand, nanotechnology has helped the development of novel biosensors for biological and medical applications. Nanobioconjugates that consists of various functional nanoparticles linked to biological molecules have been used in many areas such as diagnostics, therapeutics, sensors, and bioengineering. Detection methods based on these nanobioconjugates show increased selectivity and sensitivity as compared with many conventional assays that rely on molecular probes.

This paper reviews recent progress in the area of DNA-based artificial nanostructures. The fabrication, properties, and applications of various DNA-based nanostructures are described. New ideas and directions on DNA nanotechnology are presented. In addition, we discuss the optical and melting properties of the DNA-linked gold nanoparticles.

# 2. DNA fundamentals

DNA is the basic building block of life. Hereditary information is encoded in the chemical language of DNA and reproduced in all cells of living organisms. The double-stranded helical structure of DNA is key to its use in self-assembly applications. Each strand of the DNA is about 2 nm wide and composed of a linear chain of four possible bases (adenine, cytosine, guanine, and thymine) on a backbone of alternating sugar molecules and phosphate ions. Each unit of a phosphate, a sugar molecule, and base is called a nucleotide and is about 0.34 nm long. The specific binding through hydrogen bonds between adenine (A) and thymine (T), and cytosine (C) and guanine (G) can result in the joining of two complementary single-stranded DNA to form a double-stranded DNA. There are two hydrogen bonds between A-T pairs and three hydrogen bonds between G-C pairs. The phosphate ion carries a negative charge in the DNA molecule, which results in electrostatic repulsion of the two strands. In order to keep the two strands together, positive ions must be present in the solution to keep the negative charges neutralized. The joining of two complementary single strands of DNA through hydrogen bonding to form a double-stranded DNA is called hybridization. If a double-stranded DNA is heated above a certain temperature, the two strands will start to dehybridize and eventually separate into single strands. The center temperature of this transition is called the melting temperature, $T_m$, which is a sensitive function of environmental conditions such as ionic strength, pH, and solvent conditions. As the temperature is reduced, the two strands will eventually come together by diffusion and rehybridize to form the double-stranded structure. These properties of the DNA can be utilized in the ordering and assembly of artificial structures if these structures can be attached to DNA.

## 3. Attachment of DNA to surfaces

The first step toward DNA-based nanotechnology is to attach DNA molecules to surfaces. So far, the most widely used attachment scheme utilizes the covalent bond between sulfur and gold [1-7]. Nuzzo and Allara first reported the formation of long chain ω-substituted dialkyldisulfide molecules on a gold substrate [1]. Bain et al. [5] demonstrated a model system consisting of long-chain thiols that adsorb from solution onto gold to form densely packed, oriented monolayers. The bonding of the sulfur head group to the gold substrate is in the form of a metal thiolate, which is a very strong bond (~ 44 kcal/mol), and hence the resulting films are quite stable and very suitable for surface attachment of functional groups. For example, the DNA molecule can be functionalized with a thiol (S-H) or a disulfide (S-S) group at the 3' or 5' end. Hickman et al. also demonstrated the selective and orthogonal self-assembly of disulfide with gold and isocyanide with platinum [6]. It should be noted that there are also other strategies to attach DNA to surfaces, for example, the covalent binding of DNA oligonucleotides to a preactivated particle surface [8] and adsorption of biotinylated oligonucleotides on a particle surface coated with avidin [9, 10]. These attachment schemes have served as the fundamental base for DNA-related self-assembly of artificial nanostructures.

## 4. Self-assembly and construction of nanostructures using DNA

There has been a tremendous interest in recent years to develop concepts and approaches for self-assembled systems. While significant work continues along this direction, it has also been recognized that the exquisite molecular recognition of various natural biological materials can be used to form a complex network of potentially useful particles for a variety of optical, electronic, and sensing applications. This approach can be considered a bottom-up approach rather than the top-down approach of conventional scaling. DNA is a particularly promising candidate to serve as a construction material in nanotechnology. Despite its simplicity, the highly specific Watson-Crick hydrogen bonding allows convenient programming of artificial DNA receptor moieties. The power of DNA as a molecular tool is enhanced by automated methods and by the PCR technique to amplify any DNA sequence from microscopic to macroscopic quantities. Another attractive feature of DNA is the great mechanical rigidity of short double helices, so that they behave effectively like a rigid rod spacer between two tethered functional molecular components on both ends. Moreover, DNA displays a relatively high physicochemical stability. Finally, nature provides a complete toolbox of highly specific enzymes that enable the processing of the DNA material with atomic precision and accuracy.

## 4.1 Nanostructures of pure DNA

Seeman and co-workers [11] were the first to exploit DNA's molecular recognition properties to design complex mesoscopic structures based solely on DNA. In their work, branched DNA was used to form stick figures by properly choosing the sequence of the complementary strands. Macrocycles, DNA quadrilateral, DNA knots, Holliday junctions, and other structures were designed. Fig. 1 shows a four-armed stable branched DNA junction made by DNA molecules and the use of the branched junctions to form periodic crystals [12]. The same group also reported the design of two-dimensional crystalline forms of DNA double crossover molecules that are programmed to self-assemble by the complementary binding of the sticky ends of the DNA molecules [13]. These lattices can also serve as scaffolding material for other biological materials. A detailed review on this topic can be found in Ref. [14].

Other researchers also have put effort on using DNA to design complex architectures. Bergstrom and co-workers have designed rigid tetrahedral linkers with arylethynylaryl spacers to direct the assembly of attached oligonucleotide linker arms into novel DNA macrocycles [15]. Unlike Seeman's approach where the DNA serves as both the vertices and the edges of the assembled architectures, Bergstrom's approach utilizes rigid tetrahedral organic vertices, where the attached oligonucleotides serve as the connectors for the design of more complex architectures. In principle, a variable number of oligonucleotide arms could be attached to the core tetrahedral organic linkers, thereby allowing for the construction of different types of DNA structures.

## 4.2 DNA-based assembly of metal nanoparticles

In 1996, Mirkin and co-workers [16] first described a method of assembling colloidal gold nanoparticles into macroscopic aggregates using DNA as linking elements. As illustrated in Fig. 2, this method involved attaching noncomplementary DNA oligonucleotides to the surfaces of two batches of gold particles capped with thiol groups, which bind to gold. When another oligonucleotide which is complementary to the two grafted sequences is introduced, the nanoparticles self-assemble into aggregates. This process could also be reversed when the temperature was increased due to the melting of the DNA oligonucleotides. Because of the molecular recognition properties associated with the DNA interconnects, this strategy allows one to control interparticle distance, strength of the particle interconnects, and size and chemical identity of the particles in the targeted macroscopic structure.

In the same time, Alivisatos et al. [9] also reported DNA-based techniques to organize gold nanocrystals into spatially defined structures. In their work, gold particles were attached to either the 3' or 5' end of 19 nucleotide long single-stranded DNA molecules through the well-known thiol attachment scheme. Then, 37 nucleotide long single-stranded DNA template molecules were added to the solution containing the gold nanoparticles functionalized with single-stranded DNA. The authors showed that the nanocrystals could be assembled into dimers (parallel and antiparallel) and trimers upon hybridization of the DNA molecules with that of the template molecule. Due to the ability to choose the number of nucleotides, the gold particles can be placed at defined positions from each other as schematically shown in Fig. 3.

Based on the work of Alivisatos and co-workers, Loweth et al. [17] have studied further details of the formation of the hetero-dimeric and hetero-trimeric nonperiodic nanocluster molecules. They showed exquisite control of the placement of 5 nm and 10 nm gold nanoclusters that were derivatized with single-stranded DNA. Various schemes of hetero-dimers and hetero-trimers were designed and demonstrated with TEM images [17].

Mucic et al. [18] have made the construction of binary nanoparticle networks composed of 9 nm particles and 31 nm particles, both composed of citrate-stabilized colloidal gold. These 9 and 31 nm particles are coated with different 12-mer oligonucleotides via a thiol bond. When a third DNA sequence (24-mer), which is complementary to the oligonucleotides on both particles is added, hybridization led to the association of particles. When the ratio of 9 nm to 31 nm particles is large, a binary assembly of the nanoparticles is formed. Fig. 4 shows the scheme and a TEM image of the binary nanoparticle assembly [18].

Maeda et al. recently reported two-dimensional assembly of Au nanoparticles with a DNA network template [19]. Fig. 5 illustrates their approach. First, a gold nanoparticle is attached to a DNA1 molecule through the Au-thiol reaction. The DNA1 molecule is then hybridized with a DNA2 molecule possessing a counterbase sequence. Finally, the components are built into a DNA network consisting of DNA3 through the hybridization of DNA2 and DNA3. As a consequence, the Au particles are inserted into the DNA3 network template.

### 4.3 Construction of semiconductor particle arrays using DNA

Many strategies have been used to synthesize semiconductor particles and particle arrays. Coffer and co-workers were the first to utilize DNA as a stabilizer/template to form both CdS nanoparticles and mesoscopic aggregates from them [20]. Their original efforts were based on the use of linear duplexes of DNA in solution as a stabilizer for forming CdS nanoparticles. The initial results indicated that CdS nanoparticles could be formed from $Cd^{2+}$ and $S^{2-}$ in the presence of DNA. However, the role of DNA in the formation of the nanoparticles and the interactions between DNA and the particles after their formation were not clarified [20]. Further studies demonstrated that DNA base sequence, and more specifically the content of the base adenine, had a significant effect on the size of the CdS particles formed and their resulting photophysical properties [21].

In order to form well-defined mesoscale structures in solution, Coffer and co-workers developed a new strategy for binding a template DNA strand to a solid substrate [22]. This approach provides many possibilities for synthesizing mesoscale structures since particle composition, shape, length, and sequence of the DNA template can be controlled. The main drawback of this approach is the difficulty of forming monodispersed nanoparticle samples since the nature of the $DNA/Cd^{2+}$ interactions is poorly understood. In addition, the relative spacing and orientation of the resulting nanoparticles within the mesostructure are difficult to control, and consequently, tailoring and predicting the resulting properties of the materials is problematic [22].

Torimoto et al. tried another approach to assemble CdS nanoparticle using DNA [23]. The idea is to use the electrostatic interaction between the cationic surface modifiers on the CdS nanoparticles and the phosphate groups in DNA double strands. Fig. 6 shows the schematic illustration of their approach.

Mitchell et al. used thiolated oligonucleotides to partially displace mercaptopropionic acid molecules from the surface of the dots [24]. However, the presence of carboxyl groups on the surface of the nanocrystals was found to cause strong nonspecific binding to the oligonucleotides probe backbone. To overcome this problem, Pathal et al. developed a strategy in which hydroxylated CdSe/ZnS nanocrystals were covalently attached to oligonucleotide sequences via a carbamale linkage [25].

Tour and co-workers taken an approach similar to that of Coffer in using DNA to assemble DNA/fullerene hybrid organic materials [26]. In their strategy, the negative phosphate backbone of DNA was used as a template to bind and organize $C_{60}$ fullerene molecules modified with a N,N-dimethylpyrrolidinium iodide moiety into defined mesoscopic architectures. The modified fullerene is electrostatically complexed to the DNA backbone through cation exchange with sodium in DMSO as depicted in Fig 7.

### 4.4 DNA-directed nanowires

The concept of DNA-mediated self-assembly of nanostructures has also been extended to metallic nanowires [27-29]. Braun et al. have utilized DNA as a template to grow conducting silver nanowires [27]. The basic assembly scheme for constructing a Ag nanowire attached to two gold electrodes is outlined in Fig 8. Two gold electrodes separated by a defined distance (12-16 µm) were deposited onto a glass slide using photolithography. The gold electrodes subsequently were modified with noncomplementary hexane disulfide modified oligonucleotides through well-established thiol adsorption chemistry on Au. Subsequently, a fluorescently labeled strand of DNA containing sticky ends that are complementary to the oligonucleotides attached to the electrodes is introduced. Hybridization of the fluorescently tagged DNA molecule to the surface-confined alkylthiololigonucleotides was confirmed by fluorescence microscopy, which showed a fluorescent bridge connecting the two electrodes (see Fig. 9). After a single DNA bridge was observed, the excess hybridization reagents were removed. Silver ions then were deposited onto the DNA through cation exchange with sodium and complexation with the DNA bases. This process can be followed by monitoring the quenching of the fluorescent tag on the DNA by the Ag ions. After almost complete quenching of the fluorescence, the silver ion bound to the template DNA is reduced using standard hydroquinone reduction procedures to form small silver aggregates along the backbone of the DNA. A continuous silver wire is then formed by further Ag ion deposition onto the previously constructed silver aggregates followed by reduction. The wires are comprised of 30-50 nm Ag grains that are contiguous along the DNA backbone. Two terminal electrical measurements subsequently were performed on the Ag wire depicted in the AFM image. When the current-voltage characteristics of the Ag wire were monitored, no current was observed at near zero bias (10 V in either scan direction), indicating an extremely high resistance. At a higher bias, the wire becomes conductive. Surprisingly, the current-voltage characteristics were dependent on the direction of the scan rate, yielding different I-V curves. Although not well understood, it was postulated that the individual Ag grains that comprise the Ag nanowires may require simultaneous charging, or Ag corrosion may have occurred, resulting in the high resistance observed at low bias. By depositing more silver and thereby growing a thicker Ag nanowire, the no-current region was reduced from 10 V to 0.5 V, demonstrating crude control over the electrical properties of these systems. In addition, control experiments where one of the components (DNA or Ag) was removed from the assembly produced no current, establishing that all of the components are necessary to form the conducting Ag nanowires. This work is a proof-of-concept demonstration of how DNA can be used in a new type of chemical lithography to guide the formation of nanocircuitry.

Martin et al. also reported the fabrication of Au and Ag wires using the DNA as a template or skeleton [28]. The basic idea behind this work is to fabricate gold or platinum metal wires, functionalize these wires with exchange and formation of complexes between the gold and the DNA bases. Current voltage characteristics were measured to demonstrate the possible use of these nanowires. The authors also reported the formation of luminescent self-assembled poly (p-phenylene vinylene) wires for possible optical applications [29]. The work has a lot of potential and much room for further research to control the wire width, the contact resistances between the gold electrode and the silver wires, and use of other metals and materials.

More recently, Yan et al. [30] demonstrated the design and construction of a DNA nanostructure that has a square aspect ratio and readily self-assembles into two distinct lattice forms: nanoribbons and two-dimensional nanogrids. The $4 \times 4$ tile contains four four-arm DNA branched junctions pointing in four directions. Such nanogrids have a large cavity size, which may serve as binding or tethering sites for other molecular components. For example, the loops at the center of each $4 \times 4$ tile can be modified with appropriate functional groups and for used as a scaffold for directing periodic assembly of desired molecules. Periodic protein arrays were achieved by templated self-assembly of streptavidin onto the DNA nanogrids. The authors also used the $4 \times 4$ tile assemblies as templates to construct a highly conductive, uniform-width silver nanowire. A two-terminal I-V measurement of the resulting silver nanowire was conducted at room temperature. The I-V curve is linear, demonstrating Ohmic behavior in the range of $-0.2$ to 0.2 V. This nanowire is easily reproducible and has markedly higher conductivity than previously reported double-helix DNA-templated silver nanowires [27].

In addition to using DNA, there are also reports on using other biomolecules to fabricate nanowires. Djalali et al. recently developed a new biological approach to fabricate Au nanowires using sequenced peptide nanotubes as templates [31]. The fabrication process is illustrated in Fig. 10. Briefly, Au ions are captured by imidazole and amine groups of the sequenced peptides on the nanotubes, and then the trapped Au ions in the peptides nucleate into Au nanocrystals after reducing those ions by hydrazine hydrate. This approach has potential to control the size and the packing density of the Au nanocrystals by simply adjusting external experimental conditions such as pH, temperature, and ion concentration.

## 4.5 DNA functionalized with carbon nanotubes

Single-walled carbon nanotubes (SWNT) are composed of a single layer of graphene sheet rolled-up into a cylinder, with diameters in the range 1-2 nm [32-34]. Fig. 11 shows a TEM image of a single-walled carbon nanotube. Since the discovery of single-walled carbon nanotubes [33-35], this new class of materials has demonstrated great potential to make a major contribution to a variety of nanotechnology applications, including molecular electronics [36], hydrogen storage media [37], and scanning probe microscope tips [38]. Carbon nanotubes can be expected to provide a basis for a future generation of nanoscale devices, and it has been predicted that modification of SWNT will lead to an even more diverse range of applications. For example, the electrical properties of empty SWNT are extremely sensitive to their structure and the existence of defects, which imposes great difficulty for using unfilled nanotubes in electronic device applications. The property of filled SWNT, on the other hand, will be dominated by the filling materials, and therefore, filled nanotubes will be more robust in applications such as nanoelectronics [39-41]. The unusual physical and chemical properties also depend on the nanotube diameter and helicity. Kiang and co-workers developed an efficient method for catalytic synthesis of SWNTs with a wide range of diameters [42, 43]. Many applications would benefit from the availability of SWNT of varying diameters and helicities.

At present, SWNT-based devices are fabricated by "top-down" lithographic methods. The construction of more complex architectures with high device density requires the development of a "bottom-up," massively parallel strategy that exploits the molecular properties of SWNTs. DNA-guided assembly of carbon nanotubes could be one way toward this aim. Besides, carbon nanotubes have useful properties for various potential applications in biological devices. For instance, nanotubes can be used as electrodes for detecting biomolecules in solutions, similar to commonly used conventional carbon based electrode materials. Also, the electrical properties of SWNTs are sensitive to surface charge transfer and changes in the surrounding electrostatic environment, undergoing drastic changes simply by adsorption of certain molecules or polymers. SWNTs are therefore promising as chemical sensors for detecting molecules in the gas phase and as biosensors for probing biological processes in solutions.

One way to link DNA with nanotubes is via noncovalent interactions [44-47]. Dai and co-workers described a simple and general approach to noncovalent functionalization of the sidewalls of SWNTs and subsequent immobilization of various biological molecules onto nanotubes with a high degree of control and specificity [46]. Their method involves a bifunctional molecule, 1-pyrenebutanoic acid, succinimidyl ester, irreversibly adsorbed onto the inherently hydrophobic surfaces of SWNTs in an organic solvent dimethylformamide (DMF) or methanol. The pyrenyl group, being highly aromatic in nature, is known to interact strongly with the basal plane of graphite via $\pi$-stacking, and also found to strongly interact with the sidewalls of SWNTs in a similar manner, thus providing a fixation point for the biomolecule on the nanotubes. The anchored molecules on SWNTs are highly stable against desorption in aqueous solutions. This leads to the functionalization of SWNTs with succinimidyl ester groups that are highly reactive to nucleophilic substitution by primary and secondary amines that exist in abundance on the surface of most proteins (Fig. 12). The mechanism of protein immobilization on nanotubes, then, involves the nucleophilic substitution of N-hydroxysuccinimide by an amine group on the protein, resulting in the formation of an amide bond. This technique enables the immobilization of a wide range of biomolecules on the sidewalls of SWNTs with high specificity and efficiency.

Meanwhile, people also tried to use covalent chemistry to link DNA with nanotubes because covalent interaction is expected to provide the best stability, accessibility, and selectivity during competitive hybridization. Hamers and co-workers have developed a multi-step route to the formation of covalently linked adducts of SWNT and DNA oligonucleotides [48]. Figure 13 shows an overview of the covalent attachment process. Purified SWNTs were oxidized to form carboxylic acid groups at the ends and sidewalls. These were reacted with thionyl chloride and then ethylenediamine to produce amine-terminated sites. The amines were then reacted with the heterobifunctional cross-linker succinimidyl 4-(N-maleimidomethyl)cyclohexane-1-carboxylate, (SMCC), leaving the surface terminated with maleimide groups. Finally, thiol-terminated DNA was reacted with these groups to produce DNA-modified SWNTs. It is found that DNA molecules covalently linked to SWNTs are accessible to hybridization as evidenced by strong tendency in hybridization with molecules having complementary sequences compared with noncomplementary sequences.

Williams et al. [49] recently developed a way to couple SWNTs covalently to peptide nucleic acid (PNA, an uncharged DNA analogue [50]) and to hybridize these macromolecular wires with complementary DNA. This technique is shown in Fig. 14. It is found that DNA attachment occurs predominantly at or near the nanotube ends. Dwyer et al. also reported some progress toward the DNA-guided assembly of carbon nanotubes [51]. They used amine-terminated DNA strands to functionalize the open ends and defect sites of single-walled carbon nanotubes. Fig. 15 illustrates the DNA/nanotube reaction scheme. Nguyen et al. developed an approach for the attachment of DNA to oxidatively opened ends of multiwalled carbon nanotube arrays [52]. In addition to necessary steps such as opening the closed carbon nanotube (CNT) ends and removing metal catalyst at the nanotube tips, a novel and critical step in their approach is the deposition of a spin-on glass (SOG) film inside hydrophobic CNT arrays. It is found that the SOG improves the mechanical rigidity of the CNT array as well as enhances the DNA coupling efficiency.

In addition to DNA, there are also plenty of interests in functionalizing carbon nanotubes with other biomolecules such as protein, peptide, etc. Pantarotto et al. recently reported the synthesis, structural characterization, and immunological properties of carbon nanotubes functionalized with peptides [53]. They employed two different methods (Fig. 16) to link bioactive peptides to SWNTs through a stable bond: (i) the fragment condensation of fully protected peptides and (ii) selective chemical ligation.

## 4.6 Field effect transistor based on DNA

Mrauccio et al. [54] have demonstrated a field effect transistor based on a deoxyguanosine derivative (a DNA base). This three-terminal field effect nanodevice was fabricated starting from a deoxyguanosine derivative ($dG(C_{10})_2$). Guanosine has been chosen because of its peculiar sequence of H-bond donor or acceptor groups, and because it has the lowest oxidation potential among the DNA bases, which favors self-assembly and carrier transport, respectively. Such a guanosine supramolecular assembly has the form of long ribbons (see Figure 17), with a strong intrinsic dipole moment along the ribbon axis that causes current rectification in transport experiments. The prototype structure of this nanodevice is a planar metal-insulator-metal nanojunction, consisting of two arrow-shaped metallic electrodes facing each other and connected by the supramolecular structures. A third electrode (gate) is deposited on the back of the device to produce a field effect transistor (see Figure 17). The experiments on transport through the source and drain electrodes interconnected by self-assembled guanine ribbons suggest that these devices behave like p-channel MOSFETs. The devices exhibit a maximum voltage gain of 0.76. This prototype transistor represents a starting point toward the development of biomolecular electronic devices. In particular, the demonstration of a three-terminal field effect device (field effect transistor) consisting of source and drain contacts interconnected by a molecular layer, and a third contact (gate) to modulate the drain-source current (Ids), is a crucial step for the development of a molecular electronics road map.

### 4.7 Self-assembly using artificial DNA

Among the variety of approaches to DNA-based supramolecular chemistry, the strategy of replacing DNA natural bases by alternative bases that possess distinct shape, size, or function has allowed the modification of DNA in a highly specific and site-selective manner [55-59]. One good example is the replacement of the natural bases by artificial nucleosides or nucleoside mimics [56]. However, this approach is restricted to molecules with shapes and sizes that are commensurate to normal bases to ensure that the DNA modifications occur highly specifically and site selectively [56]. Recently, a new generation of such nucleoside mimics was reported in which the hydrogen bonding interactions were replaced by metal-mediated base pairing [60-64]. The advantage of this modification strategy is that it allows the metal ions to be replaced in the interior of the DNA duplex. This represents an important structural prerequisite for the development of new molecular devices based on interacting metal centers. Metal ions such as $Cu^{2+}$, $Pd^{2+}$, and $Ag^+$, have been successfully incorporated as artificial DNA bases into oligonucleotides by different groups (Fig. 18) [58]. Introduction of such metal-induced base pairs into DNA would not only affect the assembly-disassembly processes and the structure of DNA double strands but also confer a variety of metal-based functions upon DNA. A remarkable consequence of the insertion of just one artificial metal-ion-mediated base pair is that the thermal stability of the modified DNA duplex is strongly enhanced relative to one with normal hydrogen-bond interactions.

Tanaka et al. [60] showed that replacement of hydrogen-boned base pairing present in natural DNA by metal-mediated base pairing, with the subsequent arrangement of these metallo-base pairs into a direct stacked contact, could lead to "metallo-DNA" in which metal ions are lined up along the helix axis in a controlled and stepwise manner. Later, they successfully arranged $Cu^{2+}$ ions into a magnetic chain using the artificial DNA [65]. Fig. 19 shows the schematic representation of Cu-mediated duplex formation between two artificial DNA strands in which hydroxypyridone nucleobases replace natural base pairs. The most important structural feature of this artificial DNA is the alignment of the $Cu^{2+}$ ions along the axes inside the duplexes. The canonical helical conformation of these DNA-like duplexes ensures regular $Cu^{2+}$-$Cu^{2+}$ distances. From EPR signals the distance in the artificial DNA duplexes was estimated to be 3.7 Å, which is remarkably similar to the distance between two adjacent base pairs in natural DNA duplexes (3.4 Å).

## 5. DNA-based nanomaterials as biosensors

In recent years, there have been significant interests in using novel solid-state nanomaterials for biological and medical applications. The unique physical properties of nanoscale solids (dots or wires) in conjunction with the remarkable recognition capabilities of biomolecules could lead to miniature biological electronics and optical devices including probes and sensors. Such devices may exhibit advantages over existing technology not only in size but also in performance. In this section, we describe some good examples that utilize nanostructured materials conjugated with biomolecules as novel biosensors.

Sequence-specific DNA detection is an important topic because of its application in the disgnosis of pathogenic and genetic diseases. Many detection techniques have been developed that rely upon target hybridization with radioactive, fluorescent, chemiluminescent, or other types of labeled probes. Moreover, there are indirect detection methods that rely on enzymes to generate colorimetric, fluorescent, or chemiluminescent signals. Mirkin and co-workers [66, 67] developed a novel method for detecting polynucleotides using gold nanoparticle probes. Their method utilizes the distance- and size-dependent optical properties of aggregated Au nanoparticles functionalized with 5'-(alkanethiol)-capped oligonucleotides. Introduction of a single-stranded target oligonucleotide (30 bases) into a solution containing the appropriate probes resulted in the formation of a polymeric network of nanoparticles with a concomitant red-to-pinkish/purple color change. Hybridization was facilitated by annealing and melting of the solutions, and the denaturation of these hybrid materials showed transition temperatures over a narrow range that allowed differentiation of a variety of imperfect targets. Transfer of the hybridization mixture to a reverse-phase silica plate resulted in a blue color upon drying that could be detected visually. The unoptimized system can detect about 10 femtomoles of an oligonucleotide. This method has many desirable features including rapid detection, a colorimetric response, good selectivity, and a little or no required instrumentation. Later, the same group reported a related method in which the Au nanoparticles functionalized with 5'- and 3'-(alkanethiol)-capped oligonucletides that causes a tail-to-tail alignment of Au nanoparticle probes [67]. This new system exhibits extraordinary selectivity and provides a simple means for colorimetric, one-pot detection of a target oligonucleotide in the presence of a mixture of oligonucleotides with sequences differing by a single nucleotide.

Maxwell et al. [68] also reported that biomolecules and nanoparticles can be both structurally and functionally linked to create a new class of nanobiosensors that is able to recognize and detect specific DNA sequences and single-base mutations in a homogeneous format. The principle of this detection method is illuminated in Fig. 20. At the core of this biosensor is a 2.5 nm gold nanoparticle that functions as both a nano-scaffold and a nano-quencher. Oligonucleotide molecules labeled with a thiol group are attached at one end of the core and a fluorophore at the other end. This hybrid construct is found to spontaneously assemble into a constrained arch-like conformation on the particle surface. In the assembled state, the fluorophore is quenched by the nanoparticle. Upon target binding, the constrained conformation opens and the fluorophore leaves the surface because of the structural rigidity of the hybridized DNA (double-stranded), and fluorescence is restored. This structural change generates a fluorescence signal that is highly sensitive and specific to the target DNA.

Cao et al. [69] have also developed a nanoparticle-based method for DNA and RNA detection in which Au nanoparticle probes are labeled with oligonucleotides and Raman-active dyes. The scheme of their method is shown in Fig. 21. The gold particles form a template for silver reduction, and the silver coating acts as a surface-enhanced Raman scattering promoter for the dye-labeled particles that have been captured by target molecules and an underlying chip in microarray format. Compared with fluorescence-based chip detection, this nanoparticle-based methodology offers several advantages. The ratio of Raman intensities can be extracted from a single Raman spectrum with single-laser excitation. Second, the number of available Raman dyes is much greater than the number of available and discernable fluorescent dyes. Therefore, this method offers potentially greater flexibility, a larger pool of available and nonoverlapping probes, and higher multiplexing capabilities than do conventional fluorescence-based detection approaches.

A major challenge in the area of DNA detection is the development of methods that do not rely on polymerase chain reaction or comparable target-amplification systems that require additional instrumentation and reagents. Park et al. [70] reported an electrical method for DNA detection. They find that the binding of oligonucleotides functionalized with gold nanoparticles leads to conductivity changes associated with target-probe binding events. The binding events localize gold nanoparticles in an electrode gap; silver deposition facilitated by these nanoparticles bridges the gap and leads to readily measurable conductivity changes. With this method, they have detected target DNA at concentration as low as 500 femtomolar with a point mutation selectivity factor of ~ 100,000:1.

Wang et al. [71] reported a nanoparticle-based protocol for detecting DNA hybridization. The idea is based on a magnetically induced solid-state electrochemical stripping detection of metal tags. Their approach involves the hybridization of a target oligonucleotide to probe-coated magnetic beads, followed by binding of the streptavidin-coated gold nanoparticles to the captured target, catalytic silver precipitation on the gold particle tags, a magnetic collection of the DNA-linked particle assembly and solid-state stripping detection. The DNA hybrid bridges the metal nanoparticles to the magnetic beads, with multiple duplex links per particle. Most of the three-dimensional DNA-linked aggregate is covered with silver following the catalytic precipitation of silver on gold. Such DNA-linked particle assembly can thus be collected magnetically and anchored onto the thick-film working electrode. This leads to a direct contact of the silver with the surface and enables the solid-state electrochemical transduction.

Weizmann et al. also emploied nucleic acid-functionalized magnetic particles for amplified DNA sensing and immunosensing [72]. Fig. 22 outlines the concept for the amplified detection of a target DNA 2. Amine-functionalized borosilicate-based magnetic particles were modified with DNA 1 using the heteronifunctional corss-linker 3-maleimidopropionic acid N-hydroxysuccinimide ester. The probe 1 is complementary to a part of the target sequence 2. The 1-functionalized magnetic particles are hybridized in a single step with a mixture that includes the target 2 and the biotin-labeled nucleic acid 3 that is complementary to the free segment of 2. The three-component double-stranded DNA assembly is then interacted with avidin-horseradish peroxidase (HRP) that act as a biocatalytic label. The DNA/avidin-HPR-functionalized magnetic particles are subsequently mixed with magnetic particles modified with the naphthoquinone unit 4. The mixture of the magnetic particles is then attracted to the electrode supported by means of an external magnet. Electrochemical reduction of the naphthoquinone to the respective hydroquinone results in the catalyzed reduction of $O_2$ to $H_2O_2$. The electrogenerated $H_2O_2$ leads, in the presence of luminal 5 and the enzyme label HRP, to the generation of the chemiluminescence signal. The chemiluminescence occurs only when the target DNA 2 is in the analyzed sample. Furthermore, the light intensity relates directly to the number of recognition pairs of 1 and 2 associated with the electrode, and thus it provides a quantitative measure of the concentration of 2 in the sample. The subsequent rotation of the particles on the surface by means of the rotating external magnet results in the enhanced electrogenerated chemiluminescence, because the magnetic particle behave as rotating microelectrode, where the interaction of $O_2$ and luminal with the catalysts on the electrode is controlled by convection rather than by diffusion. Thus, the rotation of the magnetic particles yields amplified detection of DNA.

He et al. [73] recently described a new approach for ultrasensitive detection of DNA hybridization based on nanoparticle-amplified surface plasmon resonance (SPR). Use of the Au nanoparticle tags leads to a greater than 10-fold increase in angle shift, which corresponds to a more than 1000-fold improvement in sensitivity for the target oligonucleotide as compared to the unamplified binding event. This enhanced shift in SPR reflectivity is a combined result of greatly increased surface mass, high dielectric constant of Au particles, and electromagnetic coupling between Au nanoparticles and the Au film. DNA melting and digestion experiments further supported the feasibility of this approach in DNA hybridization studies. The extremely large angle shifts observed in particle-amplified SPR make it possible to conduct SPR imaging experiments on DNA arrays. The sensitivity of this technique begins to approach that of traditional fluorescence-based methods for DNA hybridization. These results illustrate the potential of particle-amplified SPR for array-based DNA analysis and ultrasensitive detection of oligonucleotides.

Su et al. [74] demonstrated a microcantilever based mechanical resonance DNA detection using gold nanoparticle-modified probes. The core idea is to measure the mass change of a microfabricated cantilever induced by DNA hybridization through the shift of the resonance frequency of the cantilever. The hybridization is reflected by the attachment of gold nanoparticles on the cantilever and then chemically amplified by gold nanoparticle-catalyzed nucleation of silver in a developing solution. The authors claim that this method can detect target DNA at a concentration of 0.05 nM or lower. When combined with stringent washing, this technique can detect a single base pair mismatched DNA strand. The cantilever is 1/100 times smaller than its macroscopic quartz crystal microbalance counterpart, and it can be mass-produced as miniaturized sensor arrays by current processing technology. Multiple DNA detection is also possible by coating multiple cantilevers with various capture DNA strands and monitoring the change in their resonance frequencies.

Nanoparticle-based biosensors can also be used for the physical and chemical manipulation of biological systems. Hamad-Schifferli et al. [75] recently demonstrated remote control of the hybridization behavior of DNA molecules by inductive coupling of a radio-frequency magnetic field to a Au nanocrystal covalently linked to DNA. When the magnetic field is on, the inductive coupling to the Au nanocrystal increases the local temperature of the bound DNA, thereby inducing denaturation while leaving surrounding molecules relatively unaffected. Removing the field restores the hybridization of DNA. Because dissolved biomocelues dissipate heat in less than 50 picoseconds, the switch is fully reversible. This concept shows promising potential for the control of hybridization and other biological functions on the molecular scale.

There are also intense studies on semiconductor quantum dots conjugated with biomolecules as novel probes [76-78]. These nanometer-sized conjugates are water-soluble and biocompatible, and provide important advantages over organic dyes and lanthanide probes. In particular, the emission wavelength of quantum-dot nanocrystals can be continuously tuned by changing the particle size, and a single light source can be used for simultaneous excitation of all different-sized dots. High-quality dots are also highly stable against photobleaching and have narrow, symmetric emission spectra. These novel optical properties render quantum dots ideal fluorophores for ultrasensitive, multicolor, and multiplexing applications in molecular biotechnology and bioengineering.

## 6.  Properties of DNA-linked gold nanoparticles

The DNA-linked gold nanoparticle assembly is a prototype of DNA-based nanostructures. The optical and melting properties of this system have attracted considerable interest because the understanding of these properties is essential for DNA-based nanotechnology [79-81]. As described in section 4, the preparation of DNA-linked gold nanoparticles involves two batches of gold particles that are functionalized with noncomplementary DNA oligonucleotides with sequences **a** and **b**, respectively. When another oligonucleotide (linker) with a complementary sequence **a'b'** is introduced, the gold nanoparticles self-assemble into aggregates. In this section, we describe the optical and melting properties of such DNA-linked gold nanoparticle assemblies.

### 6.1 Aggregation of DNA-modified gold nanoparticles

UV-visible absorption spectroscopy is a suitable tool to study the optical properties of DNA-linked gold nanoparticle aggregation because DNA bases and gold nanoparticles have strong absorption in the UV region (~ 260 nm) and the visible light region (~ 520 nm), respectively. The extinction coefficient of a collection of gold particles is sensitive to the size of the aggregates. Thus, the change in the gold extinction reflects the aggregation of the gold nanoparticles. The kinetics of aggregation of DNA-linked gold nanoparticles is studied by measuring the UV-visible spectrum as a function of time at room temperature, as illustrated in Fig. 23 [81]. Upon adding linker DNA, the DNA hybridization leads to aggregation of gold nanoparticles, as demonstrated in the gold surface plasmon peak (520 nm) shift of the DNA-modified gold nanoparticles [82, 83]. The aggregation starts with the wavelength shift of the plasmon band, followed by broadening and more shifting of the peak as hybridization continues. These results indicate that the initial aggregation takes place with increasing volume fraction, followed by increasing network size [84, 85].

### 6.2 Melting of DNA-linked gold nanoparticle aggregations

The UV-visible spectrum is also used to monitor the melting properties of the DNA-linked aggregates. The DNA double helix has a smaller extinction coefficient than does single-stranded DNA due to hypocromism, and, therefore, the absorption intensity at 260 nm increases as a result of DNA melting. Meanwhile, the melting will also cause sharp changes in the gold nanoparticle extinction coefficient due to the dissociation of the aggregate. Therefore, the melting can be monitored at either 260 nm or 520 nm as a function of temperature. The 260 and 520 nm melting curves are very similar, indicating that DNA and gold nanoparticle melting are closely correlated. Fig. 24 displays the melting curves of 10, 20, and 40 nm gold particles with linker DNA [81]. The melting transition width is about 5 K, compared to 12 K for melting of free DNA. The transition width as well as the melting temperature has been dramatically modified by the binding to gold particles. It is also clear that the melting properties are highly dependent on the gold nanoparticle size. For bigger gold particles the melting temperature is higher [81, 85].

### 6.3 Effects of external variables on the melting properties

Besides the particle size, the melting properties of DNA-linked gold nanoparticles are also strongly dependent on other variables, including the DNA density on the gold surface, the interparticle distance, and the salt concentration in the solution. Jin et al. recently performed a series of experiments to systematically study the effects of various external variables [80].

A high DNA surface density on the Au nanoparticle is expected to provide advantage in particle stabilization as well as to increase the hybridization efficiency. Experimental results show that for temperature below 70 ℃, the melting temperature is proportional to the DNA surface density when the nanoparticle and target concentrations are kept constant [80]. Also, a slight broadening of the melting transition was observed as the DNA density decreases.

The interparticle distance is another key parameter to control the melting properties. As gold nanoparticles are linked together via DNA hybridization, electromagnetic coupling between the nanoparticles result in significant damping of their surface plasmon resonances. The amount of extinction due to scattering is also influenced by the interparticle spacing. Interparticle distance also influences van der Waals and electrostatic forces between the particles, weakly affecting duplex DNA stability and hybridization/dehybridization properties. Experimental results show that the melting temperature increases with the length of the interparticle distance for temperatures below 70 ℃, and there is a linear relationship between the two [80].

The melting behavior of DNA-linked nanoparticle aggregation also depends on the salt concentration. In Jin et al's study [80], the melting temperature increased from 41 to 61.5 ℃ as the salt concentration was increased from 0.05 to 1.0 M while keeping the nanoparticle and linker DNA concentration constant. Moreover, increasing salt concentration also causes larger aggregates as evidenced by a larger absorption change during melting. It is believed that the salt brings about a screening effect that minimizes electrostatic repulsion between the DNA-DNA bases and between the nanoparticles, hence, strengthening the effect of the linker bond.

# 7. Conclusion

Due to its unique recognition capabilities, physicochemical stability, mechanical rigidity, and high precision processibility, DNA is a promising material for "biomolecular nanotechnology." The study of DNA-based nanostructures is hence an attractive topic. This review describes the utilization of DNA for preparing nanostructured materials and use of such nanostructures for biological and medical applications. Various DNA-based nanostructures, including nanostructures by DNA itself, DNA functionalized with metal and semiconductor nanoparticles, DNA-directed nanowires, and DNA-functionalized carbon nanotubes, were described. Some good examples of using DNA-based nanostructures as biosensor are also presented.

Though significant progress has been made, the study of DNA-based nanostructures is still at its early stage. The catalytic, electrical, magnetic, and electrochemical properties of such structures have not yet been systematically investigated, and they, therefore, represent new frontiers in this field. It is anticipated that new phenomena and useful structures will continue to emerge over the next few years. Advanced study in this field will not only provide valuable fundamental information about the collective physical and chemical properties of nanoparticles and DNA, but also may provide access to new and useful electronic and photonic materials applicable to the industry.

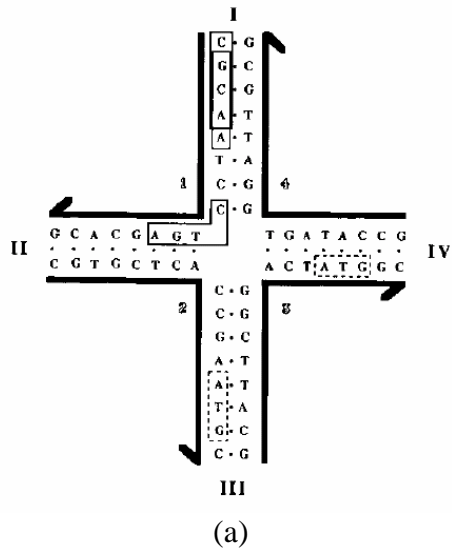

(a)

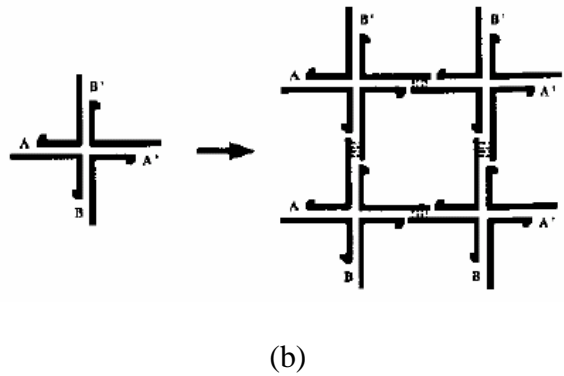

(b)

Fig. 1 (a) A four-armed stable branched DNA junction made by DNA molecules. (b) Use of the branched junctions to form periodic crystals. Reprinted from Ref. 12, N. C. Seeman, Nanotechnology 2, 149 (1991), with permission from Institute of Physics @ 1991 and Dr. Seeman.

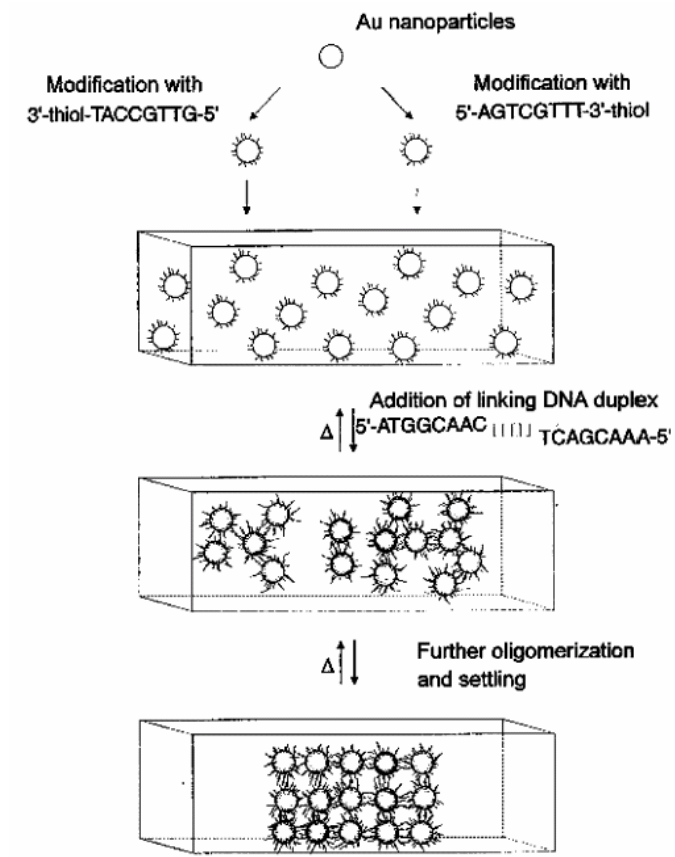

Fig. 2 Scheme showing the DNA-based gold nanoparticle assembly strategy. The scheme is not meant to imply the formation of a crystalline lattice but rather an aggregate structure that can be reversibly annealed. Reprinted from Ref. 16, C. A. Mirkin et al., Nature 382, 607 (1996), with permission from Nature Publishing Group @ 1996.

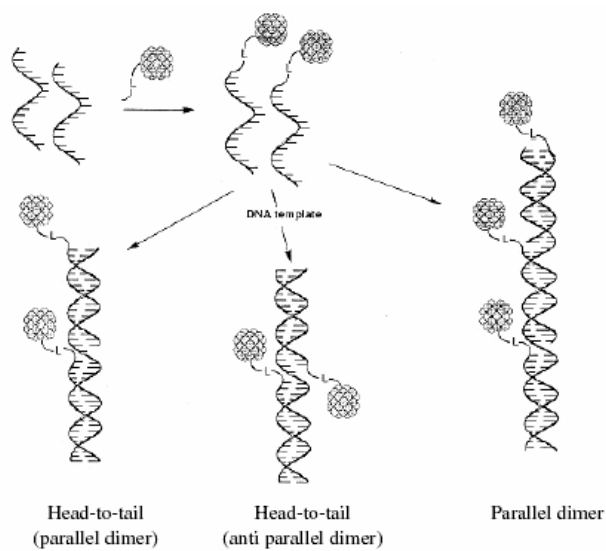

Fig. 3 Organize gold nanocrystals into spatially defined structures. Reprinted from Ref. 9, A. P. Alivisatos et al., Nature 382, 609 (1996), with permission from Nature Publishing Group @ 1996.

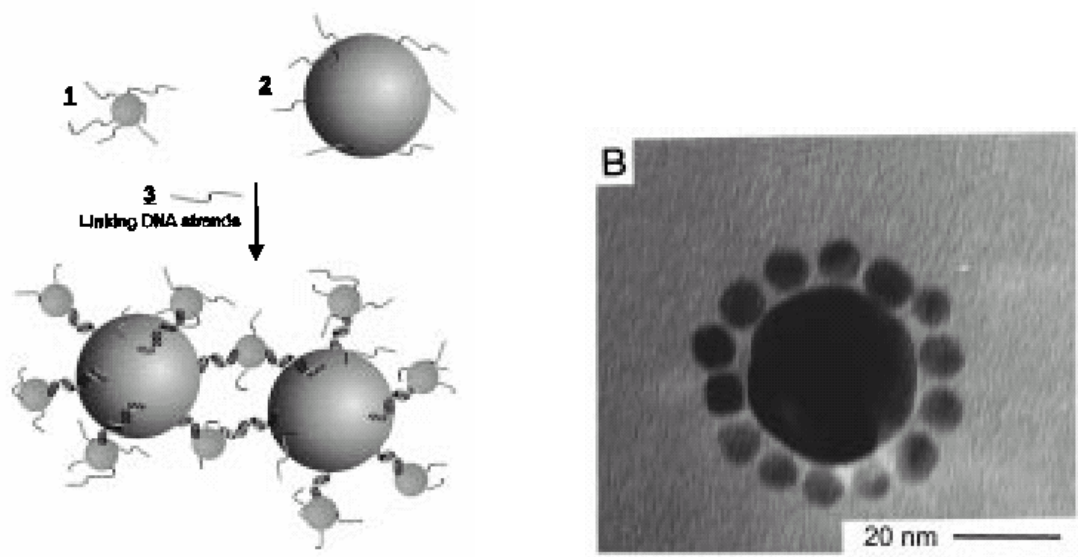

Fig. 4 DNA-directed synthesis of binary nanoparticle networks. The right panel shows a TEM image of a nanoparticle satellite structure. Reprinted from Ref. 18, R. C. Mucic et al., J. Am. Chem. Soc. 120, 12674 (1998), with permission from the American Chemical Society @ 1998.

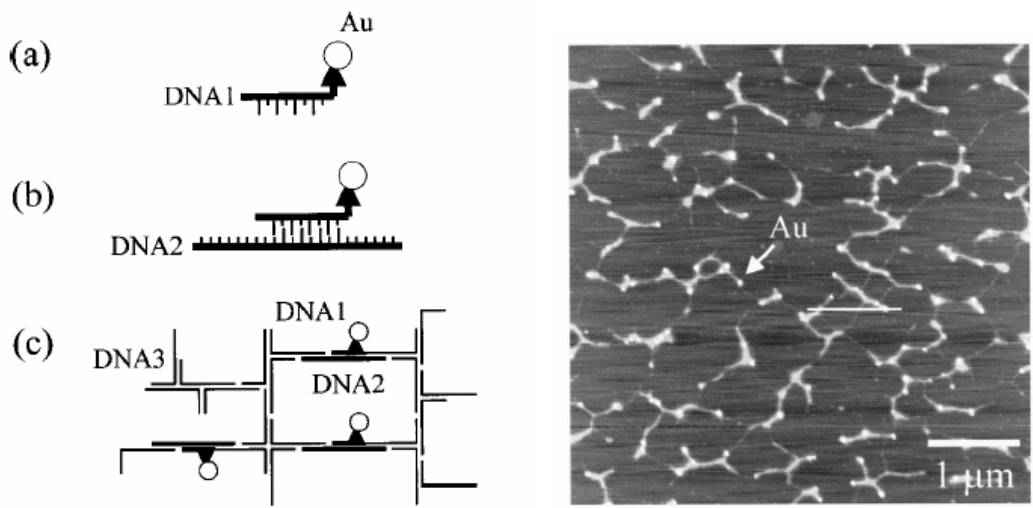

Fig. 5 (a) Approach for two-dimensional assembly of Au nanoparticles with a DNA network template. (b) AFM image of the Au-DNA network. Reprinted from Ref. 19, Y. Maeda et al., Appl. Phys. Lett. 79, 1181 (2001), with permission from American Institute of Physics @ 2001 and Dr. Kawai.

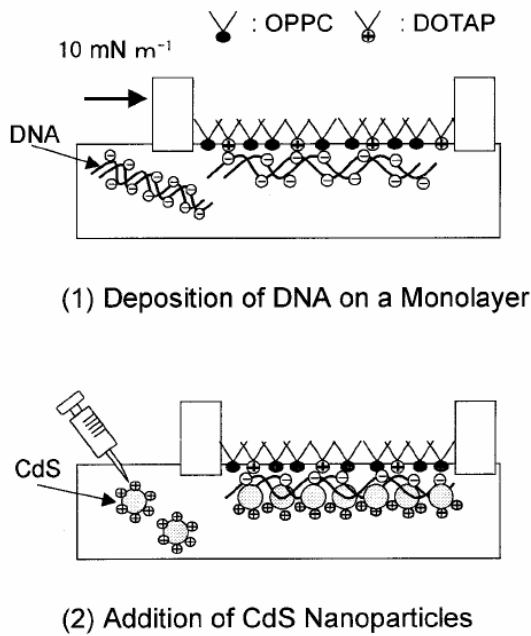

Fig. 6 Schematic illustration of the deposition of a CdS nanoparticle chain along a DNA molecule by using electrostatic interactions. Reprinted from Ref. 23, T. Torimoto et al., J. Phys. Chem. B, 103, 8799 (1999), with permission from the American Chemical Society @ 1999.

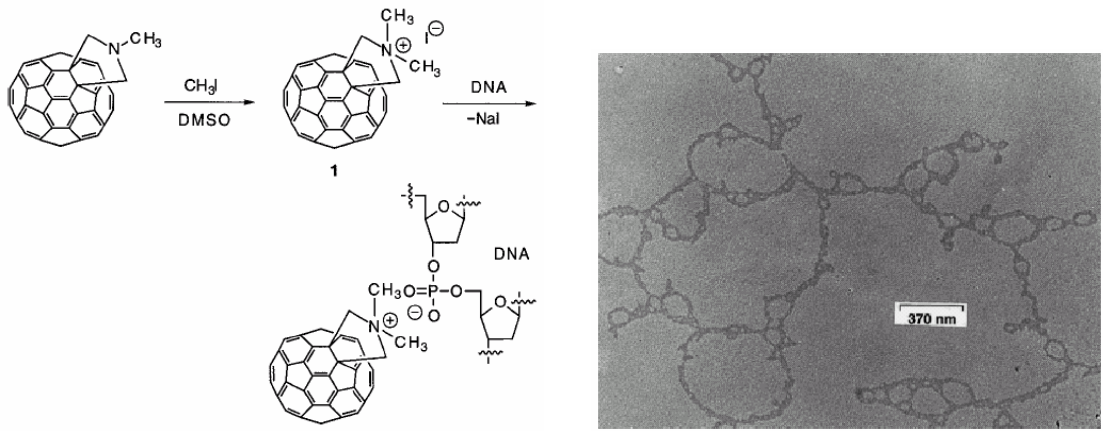

Fig. 7 Assembly of a DNA-fullerene hybrid. Reprinted from Ref. 26, A. M. Cassell et al., Angew. Chem., Int. Ed. 37, 1528 (1998), with permission from Wiley-VCH @ 1998 and Dr. Tour.

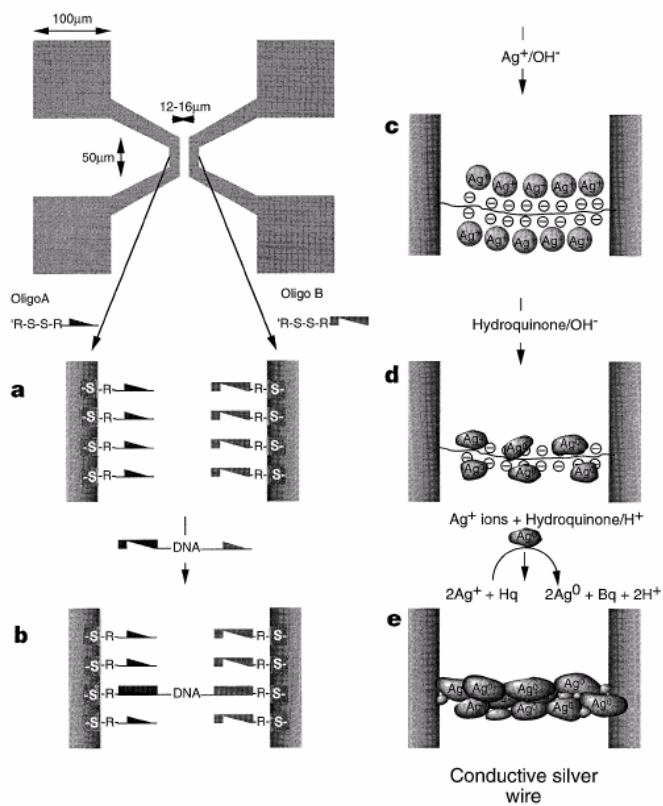

Fig. 8 DNA-templated assembly and electrode attachment of a conducting silver wire. Reprinted from Ref. 27, E. Braun et al., Nature 391, 775 (1998), with permission from Nature Publishing Group @ 1998.

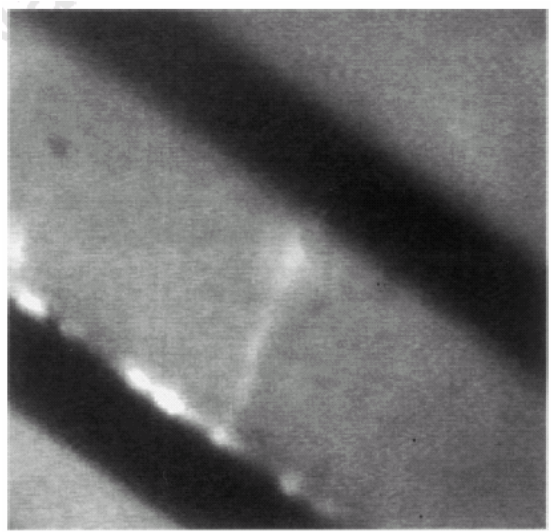

Fig. 9 Fluorescence image of a DNA bridge connecting two gold electrodes. Reprinted from Ref. 27, E. Braun et al., Nature 391, 775 (1998), with permission from Nature Publishing Group @ 1998.

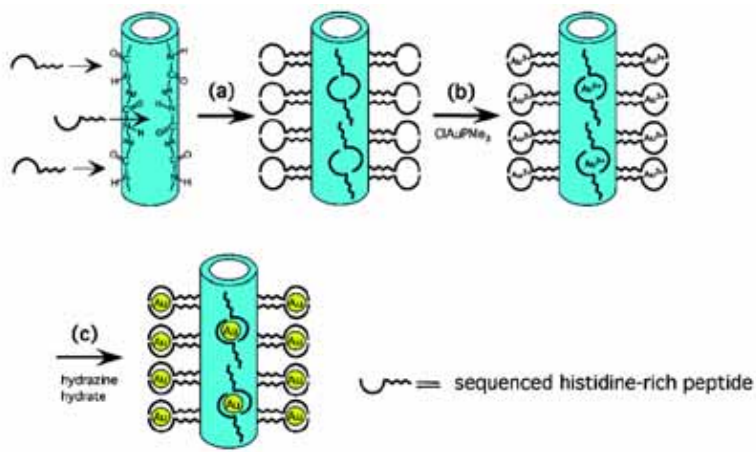

Fig. 10 Scheme for fabrication of Au nanowires using sequenced peptide nanotubes as templates. (a) Immobilization of the sequenced histidine-rich peptide at the amide binding sites of the template nanotubes. (b) Au ion immobilization on the sequenced histidine-rich peptide. (c) Au nanocrystal growth on the nanotubes nucleated at Au ion-binding sites after reducing Au ions with hydrazine hydrate. Reprinted from Ref. 31, R. Djalali, Y. Chen, and H. Matsui, J. Am. Chem. Soc. 125, 5873 (2003), with permission from the American Chemical Society @ 2003.

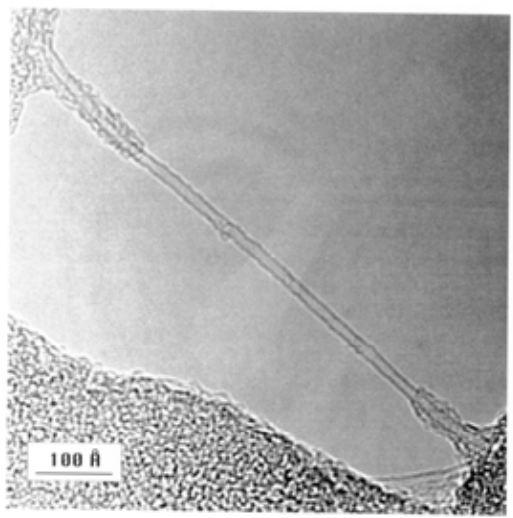

Fig. 11 TEM image of a single-walled carbon nanotube. Reprinted from Ref. 34, D. S. Bethune et al., Nature, (1993), with permission from Nature Publishing Group @ 1993.

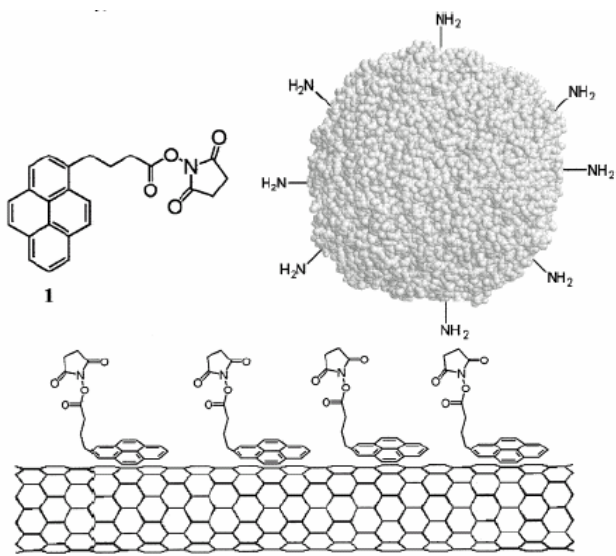

Fig. 12 Functionalization of SWNTs with succinimidyl ester groups. Reprinted from Ref. 46, R. J. Chen et al., J. Am. Chem. Soc. 123, 3838 (2001), with permission from the American Chemical Society @ 2001.

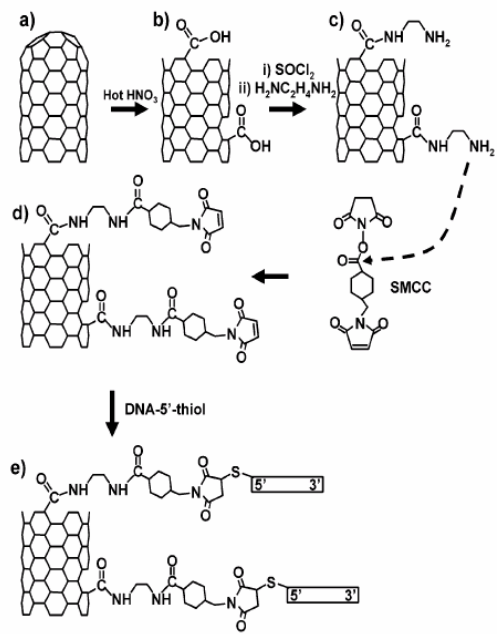

Fig. 13 An overview of the covalent attachment of DNA with SWNT. Reprinted from Ref. 48, S. E. Baker et al., Nano Lett. 2, 1413 (2002), with permission from the American Chemical Society @ 2002.

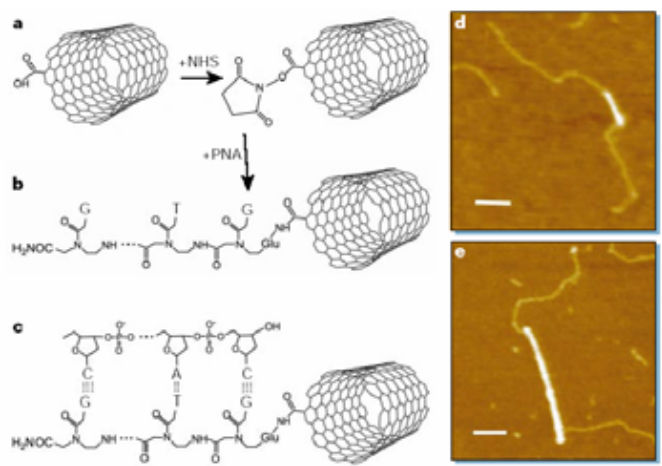

Fig. 14 Attachment of DNA to carbon nanotubes. **a, b,** N-hydroxysuccinimide (NHS) esters formed on carboxylated, SWNTs are displaced by peptide nucleic acid (PNA), forming an amide linkage. **c,** A DNA fragment with a single-stranded, 'sticky' end hybridizes by Watson–Crick base-pairing to the PNA–SWNT. **d, e,** Atomic-force microscope (TappingMode) images of PNA–SWNTs. Reprinted from Ref. 49, K. A. Williams et al., Nature 420, 761 (2002), with permission from Nature Publishing Group @ 2002.

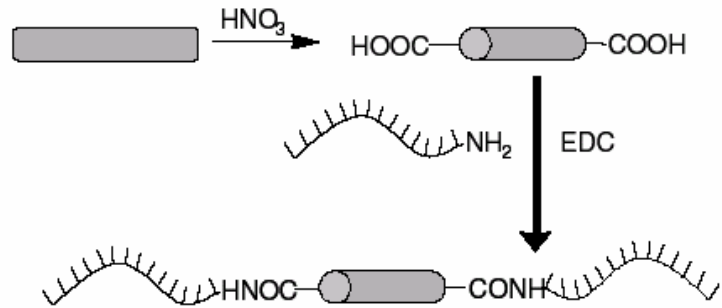

Fig. 15 The DNA/nanotube reaction scheme. Capped nanotubes are oxidatively opened and then reacted with amine-terminated single-stranded DNA. Reprinted from Ref. 51, C. Dwyer et al., Nanotechnology, 13, 601 (2002), with permission from Institute of Physics @ 2002 and Dr. Erie.

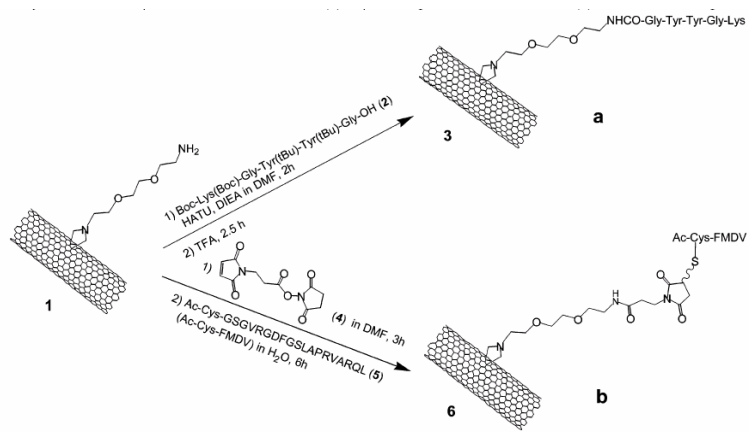

Fig. 16 Synthesis of the peptide-carbon nanotubes via (a) peptide fragment condensation and (b) chemoselective ligation. Reprinted from Ref. 53, D. Pantarotto et al., J. Am. Chem. Soc. 125, 6160 (2003), with permission from the American Chemical Society @ 2003.

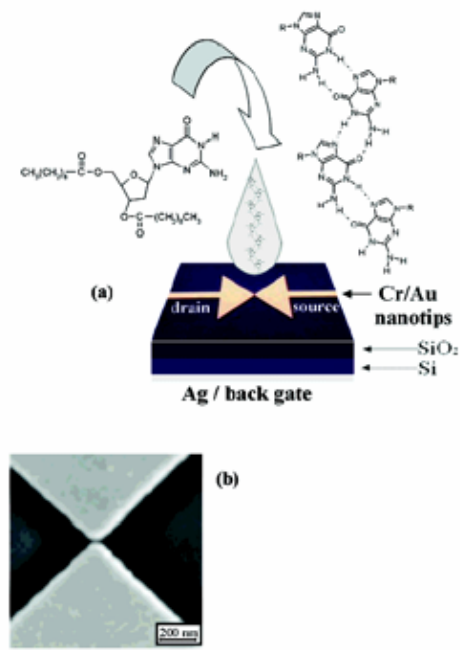

Fig. 17 A field effect transistor based on a modified DNA base. (a) Self-assembly and cast deposition of $dG(C_{10})_2$ on the three-terminal device, consisting of two arrowshaped Cr/Au (6 nm/35 nm thick) electrodes on a $SiO_2$ substrate and a third Ag back electrode (not to scale). (b) High magnification SEM image of two Cr/Au nanotips with separation of 20 nm. Reprinted from Ref. 54, G. Maruccio et al., Nano Lett. 3, 479 (2003), with permission from the American Chemical Society @ 2003.

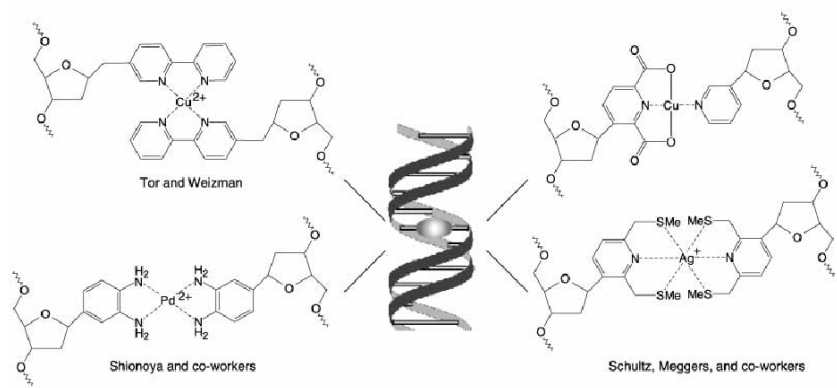

Fig. 18 Metal-ion-mediated base pairs replace natural base pairs in DNA duplexes. Reprinted from Ref. 58, H.-A. Wagenknecht, Angew. Chem. Int. Ed. 42, 3204 (2003), with permission from Wiley-VCH @ 2003.

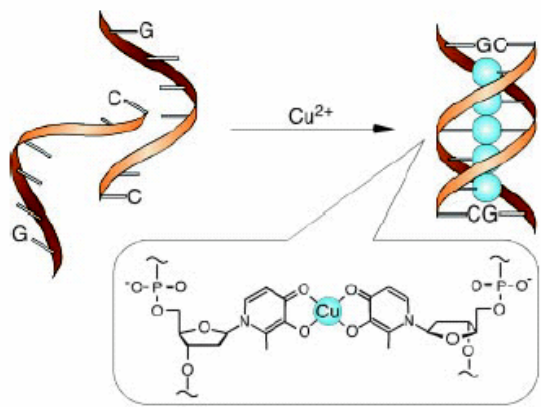

Fig. 19 Schematic representation of Cu-mediated duplex formation between two artificial DNA strands. Reprinted with permission from Ref. 65, K. Tanaka et al., Science 299, 1212 (2003). Copyright 2003 AAAS.

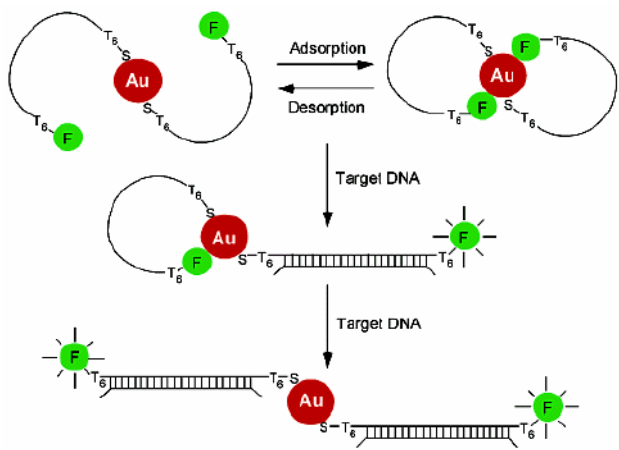

Fig. 20 Nanoparticle-based probes and their operating principles. Two oligonucleotide molecules are shown to self-assemble into a constrained conformation on each gold particle (2.5 nm diameter). A T6 spacer (six thymines) is inserted at both the 3′- and 5′-ends to reduce steric hindrance. Single-stranded DNA is represented by a single line and double-stranded DNA by a cross-linked double line. In the assembled (closed) state, the fluorophore is quenched by the nanoparticle. Upon target binding, the constrained conformation opens, the fluorophore leaves the surface because of the structural rigidity of the hybridized DNA (double-stranded), and fluorescence is restored. In the open state, the fluorophore is separated from the particle surface by about 10 nm. Au, gold particle; F, fluorophore; S, sulfur atom. Reprinted from Ref. 68, D. J. Maxwell et al., J. Am. Chem. Soc. 124, 9606 (2002), with permission from the American Chemical Society @ 2002.

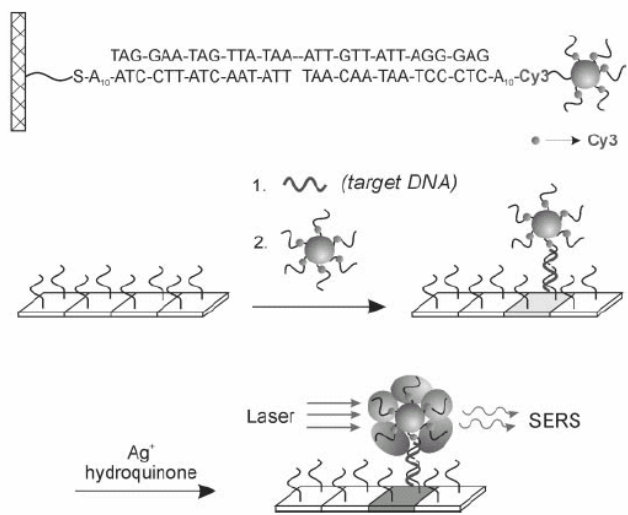

Fig. 21 Nanoparticles with Raman spectroscopic fingerprints for DNA and RNA detection. Reprinted with permission from Ref. 69, Y. Cao et al., Science 297, 1536 (2002). Copyright 2002 AAAS.

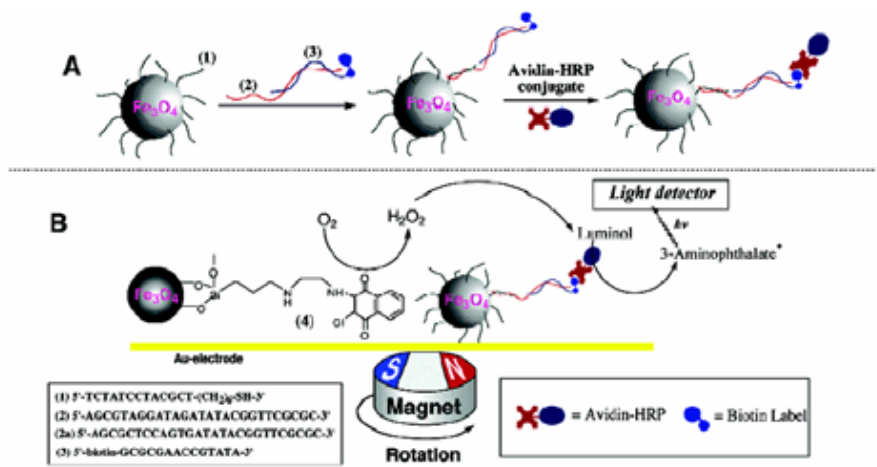

Fig. 22 (A) Preparation of DNA-functionalized magnetic particles labeled with the Avidin-HRP conjugate; (B) Amplified detection of DNA by the rotation of the labeled DNA-functionalized magnetic particles and quinone-modified magnetic particles by an external rotating magnet and the electrocatalyzed generation of chemiluminescence. Reprinted from Ref. 72, Weizmann et al., J. Am. Chem. Soc. 125, 3452 (2003), with permission from the American Chemical Society @ 2003.

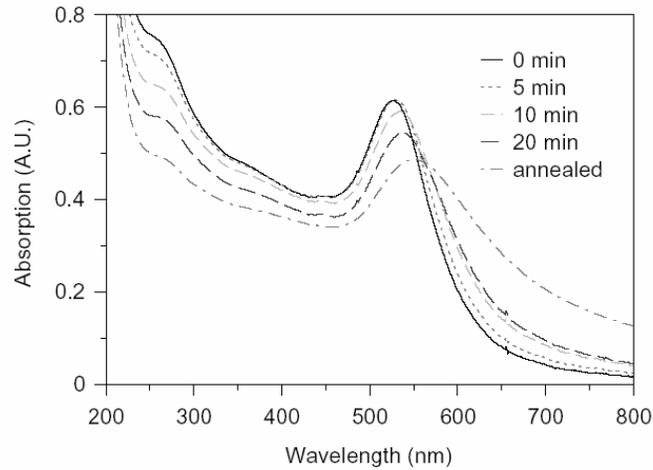

Fig. 23 Change of optical absorption spectra as a function of time of DNA-modified gold nanoparticles. Reprinted from Ref. 81, C.-H. Kiang, "Phase Transition of DNA-Linked Gold Nanoparticles," Physica A 321, 164, Copyright (2003), with permission from Elsevier.

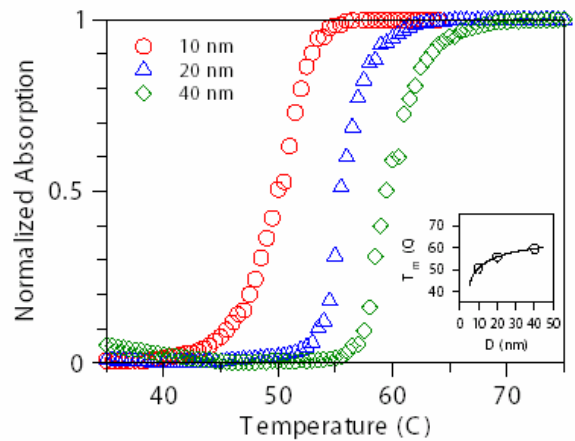

Fig. 24 Normalized melting curves of DNA-linked gold nanoparticle networks monitored at 260 nm. The inset shows the melting temperature as a function of particle diameter D. Reprinted from Ref. 81, C.-H. Kiang, "Phase Transition of DNA-Linked Gold Nanoparticles," Physica A 321, 164, Copyright (2003), with permission from Elsevier.